\documentclass[showpacs,superscriptaddress,10pt,balancelastpage,twocolumn]{revtex4}%
\usepackage[verbose,colorlinks,hyperindex,breaklinks=true,pdfusetitle,citecolor=blue,urlcolor=cyan]{hyperref}
\usepackage{amsfonts}
\usepackage{amsmath}
\usepackage{amssymb}
\usepackage{color}
\usepackage{graphicx}
\begin{document}
\title{Dynamical Localization of Bose-Einstein condensate in Optomechanics}
\author{Muhammad Ayub\footnote{\textcolor{red} {ayubok@gmail.com}}}
\affiliation{Department of Electronics, Quaid-i-Azam University, 45320, Islamabad, Pakistan.}
\affiliation{LINAC Project, PINSTECH, Nilore, 45650, Islamabad, Pakistan.}
\author{Kashif Ammar Yasir}
\affiliation{Department of Electronics, Quaid-i-Azam University, 45320, Islamabad, Pakistan.}
\author{Farhan Saif}
\affiliation{Department of Electronics, Quaid-i-Azam University, 45320, Islamabad, Pakistan.}

\begin{abstract}
We explain dynamical localization of Bose-Einstein condensate (BEC) in optomechanics both in position and 
in momentum space. The experimentally realizable optomechanical system is a Fabry-P\'{e}rot cavity with one moving 
end mirror and driven by single mode standing field. 
In our study we analyze variations in modulation strength and effective Plank's constants. 
Keeping in view present day experimental advances we provide set of parameters to observe the phenomenon 
in laboratory.
\end{abstract}
 \pacs{72.15.Rn, 03.75.Kk, 42.50.Pq}
\maketitle

\section{Introduction}

Cavity-optomechanics deals with the interaction of an optical field in a resonator with confining 
mirrors~\cite{Kippenberg}. Recent experimental advances make it possible to couple cold atoms and 
Bose-Einstein condensates (BEC), mechanical membrane and nano-sphere with the optomechanical systems
\cite{Esslinger}. Hence, these hybrid-opto mechanical systems are playground to study phenomena related 
to mirror-field interaction, and atom-field interaction which provide founding principles to develop 
numerous sensors, and devices in quantum metrology. In opto-mechanics the mechanical effects of light 
lead to cool the motion of a movable mirror to its quantum mechanical ground 
state \cite{CornnellNat2010,TeufelNat2011,CannNat2011},
and to observe strong coupling effects~\cite{GroeblacherNat2009,ToefelNat2011a,VerhagenNat2012}. 
In earlier work optomechanical systems were suggested to develop gravitational wave detectors~\cite{Braginsky} 
and to measure displacement with large accuracy~\cite{Rugar}, and in recent research work these systems lead to 
develop optomechanical crystals~\cite{EchenfieldNat2012}.  
Recent theoretical discussions and simulations on bistable behavior of BEC-optomechanical system
\cite{Meystre2010}, high fidelity state transfer \cite{YingPRL2012,SinghPRL2012}, steady-state entanglement of 
BEC and moving end mirror~\cite{Asjad2011}, macroscopic tunneling of an optomechanical membrane  
\cite{Buchmann2012} and role reversal between matter-wave and quantized light field are guiding 
towards new avenues in cavity optomechanics.  
In this paper, we discuss the dynamical localization of ultra-cold atoms or dilute BEC trapped inside a 
single mode driven optomechanical cavity  
with one fixed mirror and the other as a movable mirror. Dynamical localization is an interesting 
phenomenon in periodically driven nonlinear systems 
\cite{Fishman1982,Blumel1991,Moore1994,Bardroff1995,Schelle2009} 
and provides quantum mechanical limits on classical diffusion of wave packet.
We show that for different modulation regimes the 
dynamical localization is observed both in 
position and in momentum space. The phenomenon of dynamical localization emerges as the hybrid optomechanical 
system is 
explicitly time dependent, hence, the single mode laser field provides spatially periodic potential to BEC with 
phase modulation\cite{nature1988} due to modulated end mirror. 

In the paper, the model of the system is presented in section~\ref{sec:Model}. 
In section~\ref{sec:Langevin} we derive the Langevin equation and in section~\ref{sec:NumDL}, 
we explain dynamical localization of the condensates in the system. In section~\ref{sec:ModEffect}, 
we explain dynamical localization as a function of modulation amplitudes. Results are summarized in 
section-\ref{sec:Discussion}.

\section {The Model}\label{sec:Model}

We consider a Fabry-P\'{e}rot cavity of length $L$ with a moving end mirror driven by a single mode 
optical field of frequency $\omega_{p}$ and BEC with N-two level atoms trapped in an optical lattice potential 
\cite{Nature2008,Science2008}. Moving end mirror has harmonic oscillations with frequency $\omega_{m}$ and 
exhibits Brownian motion in the absence of coupling with radiation pressure.

The Hamiltonian of BEC-optomechanical system is,
\begin{equation}
\hat{H}=\hat{H}_{m}+\hat{H}_{a}+\hat{H}_{T}\label{1},
\end{equation}
where, $\hat{H}_{m}$ describes the intra-cavity field and its coupling to the moving end mirror, $\hat{H_{a}}$ 
accounts 
for the BEC and its coupling with intra-cavity field while, $\hat{H}_{T }$ accounts for noises and 
damping associated with the system. The Hamiltonian $\hat{H}_{m}$ is given as \cite{LawPRA1995},
\begin{equation}
\hat{H}_{m}=\hbar\triangle_{c}\hat{c}^{\dag}\hat{c}+\frac{\hbar\omega_{m}}{2}(\hat{p}^{2}+
\hat{q}^{2})-\xi\hbar\hat{c}^{\dag}\hat{c}\hat{q}-i\hbar\eta(\hat{c}
-\hat{c}^{\dag}),
\end{equation}
where, first term is free energy of the field, $\Delta_{c}=\omega_{c}-\omega_{p}$ is detuning, here, 
$\omega_{c}$ is cavity frequency and $\hat{c}^{\dag}$ ($\hat{c}$) are creation (annihilation) operators 
for intra-cavity field interacting with mirror and condensates and their commutation relation is  
$[\hat{c},\hat{c}^{\dag}]=1$. Second term represents energy of moving end mirror. Here $\hat{q}$ and $\hat{p}$ 
are dimensionless position and momentum operators for moving end mirror, such that, $[\hat{q},\hat{p}]=i$. 
Intra-cavity 
field couple BEC and moving end mirror through radiation pressure. Third term represents such coupled energy 
of moving end mirror with field and $\xi=\sqrt{2}(\omega_{c}/L)x_{0}$ is the coupling strength where, 
$x_{0}=\sqrt{\hbar/2m\omega_{m}}$, is zero point motion of mechanical mirror having mass $m$. 
Last term gives relation of intra-cavity field and output power $\vert\eta\vert=\sqrt{P\kappa/\hbar\omega_{p}}$, 
where, $P$ is the input laser power and $\kappa$ is cavity decay rate associated with outgoing modes.

Now the Hamiltonian for BEC and intra-cavity field and their coupling is derived by considering quantized motion 
of atoms along the cavity axis in one dimensional model. We assume that BEC is dilute enough and many body 
interaction effects are ignored. We have
\begin{equation}
\hat{H}_a=\int\hat{\psi}^{\dag}(x)\left(-\frac{\hslash d^{2}}{2m_{a}dx^{2}}+
\hslash U_{0}\hat{c}^{\dag}\hat{c}\cos^{2}kx\right)\hat{\psi}(x)dx,  
\end{equation}
here, $\hat{\psi}(\hat{\psi}^{\dag})$ is annihilation (creation) operator for bosonic particles and 
$U_{0}=g^2_{0}/\Delta_{a}$ is the far off-resonant vacuum Rabi frequency. Here, $\Delta_{a}$ is far-off 
detuning between 
field frequency and $\omega_{0}$, atomic 
transition frequency, $m_a$ is mass of an atom, and $k$ is the wave number. 
Due to field interaction with BEC, photon recoil takes place that generates symmetric momentum $\pm2l\hbar k$ side modes, where, $l$ is an integer. 
We assume that field is weak enough which causes low photon coupling, therefore only $0^{th}$ and $1^{st}$ 
order modes are excited while, higher order modes are ignored. 
Now $\hat{\psi}$ is depending upon these two modes~\cite{Meystre2010} and defined as, 
$\hat{\psi}(x)=\hat{a}+\sqrt{2}cos(2kx)\hat{b}/L$. Where, $\hat{a}$ and $\hat{b}$ are annihilation 
operators for $0^{th}$ and $\pm2\hslash k^{th}$ modes respectively and are related 
as $\hat{a}^{\dag}\hat{a}+\hat{b}^{\dag}\hat{b}=N$, where, $N$ is the number of bosonic particles. 
As population in $0^{th}$ mode is much larger than the population in $1^{st}$ order side mode, 
we write $\hat{a}^{\dag}\hat{a}\simeq N$ or $\hat{a}$ and $\hat{a}^{\dag}\rightarrow \sqrt{N}$. 
This is possible when side modes are weak enough and for that matter can be ignored.

By using $\hat{\psi}(x)$ defined above in Hamiltonian, we write the Hamiltonian governing the field-condensate 
interaction as,
\begin{equation}\label{Ha}
\hat{H}_{a}=\frac{\hbar U_{0}N}{2}\hat{c}^{\dag}\hat{c}+\frac{\hbar\Omega}{2}(\hat{P}^{2}
+\hat{Q}^{2})+\xi_{sm}\hbar\hat{c}^{\dag}\hat{c}\hat{Q}
\end{equation}
here, first term describes energy of field due to the condensate. We also assume large atom-field 
detuning $\Delta_{a}$, therefore, 
excited atomic levels are adiabatically eliminated. Second term expresses the energy of the condensate in the 
cavity following harmonic motion. Here, $\hat{P}=\frac{i}{\sqrt{2}}(\hat{b}-\hat{b}^{\dag})$ and 
$\hat{Q}=\frac{1}{\sqrt{2}}(\hat{b}+\hat{b}^{\dag})$ are dimensionless momentum and position operators for 
condensate which are related as $[\hat{Q},\hat{P}]=i$ and $\Omega=4\omega_{r}$ where, 
$\omega_{r}=\hbar k^{2}/2m_{a}$ is recoil frequency of an atom. 
Last term in eq. (\ref{Ha}) describes coupled energy of field and condensate 
with coupling strength $\xi_{sm}=\frac{\omega_{c}}{L}\sqrt{\hbar/m_{sm}4\omega_{r}}$, where, 
$m_{sm}=\hslash\omega_{c}^{2}/(L^{2}NU^2_{0}\omega_{r})$ is the effective mass of side mode.
  
\section{Langevin Equations}\label{sec:Langevin}

The Hamiltonian $\hat{H}_{T}$ accounts for the effects of dissipation in the intra-cavity field, damping of 
moving end mirror and depletion of BEC in the system via standard quantum noise operators ~\cite{Noise1991}. 
The total Hamiltonian $H$ leads to coupled quantum 
Langevin equations for position and momentum of moving end mirror and BEC, viz., 
\begin{eqnarray}\label{2}
\frac{d\hat{c}}{dt}&=&\dot{\hat c}=(i\tilde{\Delta}+i\xi
\hat{q}-i\xi_{sm}-\kappa)\hat{c}+\eta+\sqrt{2\kappa a_{in}},\notag \\
\frac{d\hat{p}}{dt}&=&\dot{\hat p}=-\omega_{m}\hat{q}-\xi\hat{c}^{\dag}\hat{c}
-\gamma_{m}\hat{p}+\hat{f}_{B},\notag \\
\frac{d\hat{q}}{dt}&=&\dot{\hat q}=\omega_{m}\hat{p},\notag \\
\frac{d\hat{P}}{dt}&=&\dot{\hat P}=-4\omega_{r}\hat{Q}-\xi_{sm}\hat{c}^{\dag}\hat{c}
-\gamma_{sm}\hat{P}+\hat{f}_{1m},\notag \\
\frac{d\hat{Q}}{dt}&=&\dot{\hat Q}=4\omega_{r}\hat{P}-\gamma_{sm}\hat{Q}+\hat{f}_{2m}.
\end{eqnarray}
In above equations $\tilde{\Delta}=\Delta _{c}-NU_{0}/2$, whereas $\hat{a}_{\mathrm{in}}$ is Markovian input 
noise of 
the cavity field. The term $\gamma _{m}$ gives mechanical energy decay rate of the moving end mirror and 
$\hat{f}_{B}$ is Brownian noise operator~\cite{Pater06}. The term $\gamma_{\rm sm}$ represents damping of 
BEC due to harmonic trapping potential which effects momentum side modes while, $\hat{f}_{1M}$ and $\hat{f}_{2M}$ 
are the associated noise operators assumed to be Markovian.

The coupled equations of motion which govern the dynamics of the moving end-mirror and the BEC are quantum 
Langevin equations~(\ref{2}). In adiabatic approximation we consider no thermal excitation, for the reason, we 
ignore noises and radiation effects resulting from optical damping~\cite{Kipp07,Science2008,Kippenberg}. 
Therefore, in short time limit we ignore the mechanical damping, as a result the coupled equations of motion 
for the moving end-mirror and BEC reveal their coupled nonlinear oscillator dynamics, that is,  
\begin{eqnarray}
\frac{d^{2}\hat{q}}{dt^{2}}&=&-\omega_{m}^{2}\hat{q}
 +\frac{\omega_{m}\xi\eta^{2}}{\kappa^{2}
+(\tilde{\triangle}+\xi\hat{q}-\xi_{sm}\hat{Q})^{2}}\label{3},\\
\frac{d^{2}\hat{Q}}{dt^{2}}&=&-4\omega_{r}^{2}\hat{Q}
-\frac{4\omega_{r}\xi_{sm}\eta^{2}}{\kappa^{2}
+(\tilde{\triangle}+\xi\hat{q}-\xi_{sm}\hat{Q})^{2}}\label{3}.
\end{eqnarray}
The corresponding effective Hamiltonian is described as 
$\hat{H}_{eff}=\hat{K}+\hat{V}$, where, 
$\hat{K}=\hat{p}^{2}/2$, 
here, $k^{\hspace{-2.1mm}-}=1$ is dimensionless Planck's constant. 
The potential $\hat{V}$ is governed by Eq.~(\ref{3}). We assume that for a weak coupling the moving end mirror 
behaves like 
harmonic oscillator with frequency $\omega_{m}$ and $q=q_{0}\cos(\omega_{m}t)$, where, $q_{0}$ is the maximum 
displacement from mean position. We introduce some dimensionless parameters defined as, 
$\gamma=4\omega_{r}/\omega_{m}$, $\beta=\eta^{2}/\kappa^{2}$, 
$\mu=\Delta/\kappa$, $\mu_{1}=\xi_{sm}/\kappa$, $\lambda=\frac{\xi_{sm}}{\xi}q_{0}$ and dimensionless time, 
$4\tau=\omega_{m}t$. 
Hence, from Eq.~(\ref{3}) we get,
\begin{equation}\label{4}
\hat{V}=\frac{d\hat{Q}}{d\tau}=\frac{1}{2}\hat{Q}^{2}
+\int\frac{4\gamma\xi\beta/\omega_{m}}{1+[\mu-\mu_{1}\{\hat{Q}-\lambda\cos(4\tau)\}]^{2}}dQ.
\end{equation}
Now, we write the Hamiltonian as,
\begin{equation}\label{5}
\hat{H}_{eff}=-\frac{1}{2}\frac{\partial^{2}}{\partial x^{2}}+\frac{1}{2}\hat{Q}^{2}
+\int\frac{4\gamma\xi\beta/\omega_{m}}{1+[\mu-\mu_{1}\{\hat{Q}
-\lambda\cos(4\tau)\}]^{2}}dQ.
\end{equation}
By using transformation $\hat X=\hat{Q}-\lambda\cos(4\tau)$, we find an
effective Hamiltonian for the condensate, that is,
\begin{equation}\label{6}
\hat{H}_{eff}=\frac{1}{2}\hat{\tilde{P}}^{2}+\frac{1}{2}\hat{X}^{2}+\hat{X}\lambda_{eff}\cos(4\tau)
-\frac{\gamma_{m}\beta}{\mu_{1}}\arctan(\mu-\mu_{1}\hat{X}),
\end{equation}
where, $\gamma_{m}=\frac{4\xi}{\gamma\omega_{m}}$ and $\lambda_{eff}=(1+\frac{32}{\gamma^{2}})\lambda$ is the effective modulation.

In our later work we consider the power of external field $P=0.0164mW$ with frequency 
$\omega_{p}=3.8\times2\pi\times10^{14}Hz$ and wave length $\lambda_{p}=780nm$. 
Coupling of external field and intra-cavity field is $\eta=18.4\times2\pi MHz$ and frequency of intra-cavity 
field is considered 
$\omega_{c}=15.3\times2\pi\times10^{14}Hz$ 
which produces recoil of $\omega_{r}=3.8\times2\pi kHz$ in atoms placed in cavity of length 
$l=1.25\times10^{-4}m$ and having decay rate $\kappa=1.3\times2\pi MHz$. 
Moving end mirror of cavity should be perfect reflector that oscillates with a 
frequency $\omega_{m}=15.2\times2\pi kHz$. 
The  mirror-field and condensate-field coupling strengths are, respectively, $\xi=14.39 KHz$ and 
$\xi_{sm}=15.07 MHz$. 
Detuning of the system is taken as $\Delta=\Delta_{c}+ \frac{U_{0}N}{2}=0.52\times2\pi MHz$, 
where vacuum Rabi frequency 
of the system is $U_{0}=3.1\times2\pi MHz$ and number of ultra cold atoms placed in the 
BEC-optomechanical system are $N=2.8\times10^{4}$~\cite{Esslinger,Ferdinand,AIP,Carmon05,Carmon07}.

\begin{figure*}[tp]
\centering
\includegraphics[width=16cm]{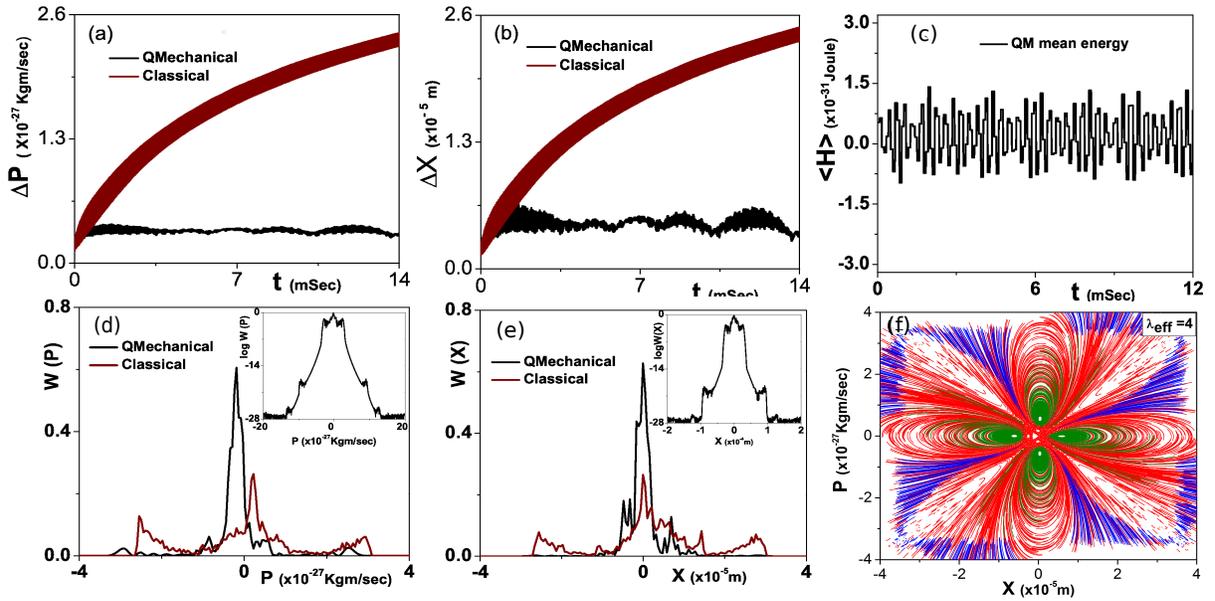}
\caption{(a) shows dispersion $\Delta p$ in momentum space and (b) shows dispersion $\Delta x$ in position space 
vs time for fixed modulation 
and gives comparison in quantum mechanical and classical behavior. 
(c) Mean energy of the condensate as a function of time.
(d) Classical and 
quantum mechanical probability distribution in momentum space. (e) Classical and quantum mechanical 
probability distribution in position space of condensate for fixed time $t=4.19\times10^{-2}sec$. 
(f) Phase Space have initial distribution from -4 to 4 both in position space $P\times10^{-26}kgm/sec$ 
and momentum space $X\times10^{-5}m$. In all these figures 
modulation $\lambda_{eff}=1.05\times10^{-5}m$. Moreover, $\mu=-0.4$, $\mu_{1}=2$, $\beta=1.8$ 
and $\gamma_{m}=0.6034$.
}
\label{b}
\end{figure*}

\section {Dynamical Localization of the condensate}\label{sec:NumDL}

The quantum dynamics of dilute BEC is explored by evolving a Gaussian material wave packet, $\Psi(X)$, 
defined at $\tau=0$ 
with, $\Delta X$ as its initial dispersion in the position space and $\Delta P$ corresponding to its initial 
dispersion in the momentum space. 

Dynamics of the condensate is shown in Fig-\ref{b}. Here, Fig-\ref{b}(a) shows classical and quantum dispersion 
in momentum space as a function of time. 
Initially, the classical and quantum dispersion in momentum space follows the classical diffusion law but 
after quantum break time, quantum dispersion of the material wave packet saturates and oscillates 
around a mean value   
while, classical dispersion increases continuously with time, following $t^{\alpha}$ law, where $\alpha=0.61$ 
showing anomalous diffusion in classical domain. Similarly, from Fig-\ref{b}(b), we note that quantum dispersion 
in position space  
saturates after quantum break time and fluctuates around average value while, classical dispersion 
in position space that shows anomalous diffusion as well. 

Here, Fig.-\ref{b}(d,e) show classical and quantum time averaged probability distributions of BEC both in momentum 
and in position coordinates, respectively, for modulation amplitude $\lambda_{eff}=1.05\times10^{-5}m$.
We calculate probabilities in momentum coordinate $W(P)=|\Psi(P)|^{2}$ and position coordinate
$W(X)=|\Psi(X)|^{2}$ after an evolution time $t=4.19\times10^{-2}sec$. 
From the numerical results, we observe that the classical probability distribution in momentum and in position 
space are 
broader than their quantum mechanical counterpart. In contrast quantum mechanical 
probability distributions are maximum at the points where the matter wave was  
initially placed and as we traverse the regions away from the initial mean values the probability of finding 
the matter wave 
both in momentum and in position space decreases.
The mixed phase space with chaotic and regular regions appear due to the presence of modulated end-mirror that 
modifies the evolution of the condensate in effective nonlinear potential seen by the condensate in the cavity, 
as shown in Fig-\ref{b}(f). 

As probability distributions are marginal distributions of phase space, in quantum mechanical probability 
distributions the peak positions both in momentum and in position coordinates correspond to the underlying 
nonlinear resonances in the Poincare surface of section. Maximum fraction of the condensate is trapped with 
momentum in the range
$-3\times10^{-27}$ to $3\times10^{-27}kg~m/sec$ 
and in the position space $-3\times10^{-5}m$ to $3\times10^{-5}m$ because of regular 
regions in this domain as shown in Fig-\ref{b}(f). 
In momentum distribution, peak at $P\sim 1.4\times10^{-30}kg~m/sec$ corresponds to 
the regular region at same momentum value. 
Similarly, in position space distribution peaks at $X\sim -6.2\times10^{-6}$ and $X\sim 6.5\times10^{-6}$ 
corresponds to the regular regions appearing in phase space. 

The part of the wave packet which is not in 
regular region experiences exponential decay which indicates that the probability of finding the condensate
decreases exponentially in the regions away from the classical resonances.
Maximum probability of finding the condensate is near the primary resonance as it originates initially from there. 
The exponential decay of distribution is one of the fundamental evidence of 
dynamical localization. 

Fig-\ref{b}(c) shows mean quantum mechanical energy 
$<H>$ of the condensate as a function of time for $\lambda_{eff}=1.05\times10^{-5}m$. 
The mean energy of the condensate shows very small fluctuation around a constant value and shows 
quantum recurrences as a function of time. 

\begin{figure}[htp]
\includegraphics[width=9cm]{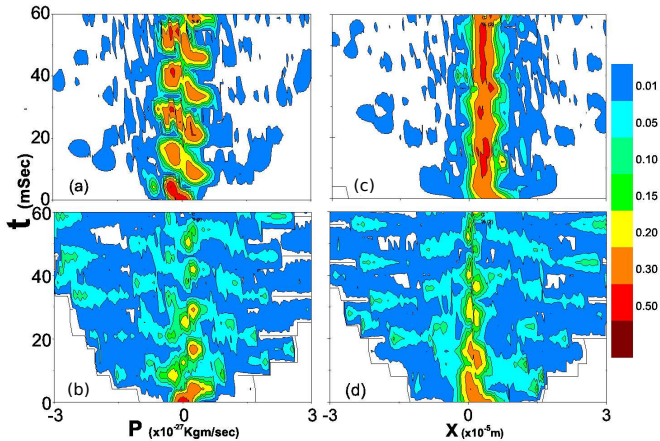}
\caption{Classical and quantum mechanical probability distribution versus time. 
(a) Quantum mechanical distribution in momentum space and (b) in position space.
(c) Classical distribution in momentum space and (d) in position space.  
The modulation is fixed as $\lambda_{eff}=1.05\times10^{-5}m$. 
The remaining parameters are same as in Fig-\ref{b}.}
\label{fig:DistVsTime}
\end{figure}

The consequence of dynamical localization in position and momentum space becomes prominent as we study 
the spatio-temporal dynamics of the condensate, presented in Fig-\ref{fig:DistVsTime}. 
We show space-time dynamics in momentum space both in quantum mechanical and classical space, respectively, 
in Fig.~\ref{fig:DistVsTime}(a) and \ref{fig:DistVsTime}(b). 
Whereas, we plot space-time dynamics in position space both quantum mechanically and classically, 
respectively, in Fig.~\ref{fig:DistVsTime}(c) and \ref{fig:DistVsTime}(d).  As the time evolves classical distributions both 
in momentum space and in position space melt down and spread over 
the entire available space with maximum probability in the resonances. On the other hand, quantum mechanical 
probability 
distributions in momentum space and in position space initially follow classical behavior, however, beyond 
quantum break time localization limits the spread. The maximum probability of finding the material wave packet 
stays around the nonlinear resonances.
The material wave packet tunnels to other resonances which provide the exponential decay as shown in 
Fig.~\ref{b}(d,e), therefore probability of finding the condensate in neighboring resonances increases with time. 

\section {Effects of modulation}\label{sec:ModEffect}

\begin{figure}[tp]
\includegraphics[width=8.5cm]{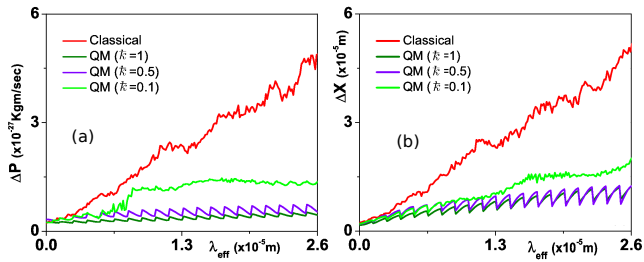}
\caption{Left figure shows classical and quantum dispersion in momentum space and right figure shows classical 
and quantum
dispersion in position space as function of modulation. These dispersion in momentum and position space are 
observed after an evolution time $t=2.09\times10^{-2}sec$. 
The other parameters are same as in Fig.~\ref{b}.}
\label{e}
\end{figure}

We numerically study the classical dispersion and quantum dispersion of the condensate both in momentum space 
Fig-\ref{e}(a) and in position space Fig-\ref{e}(b) as a function of modulation strength for a fixed time.
For a small value of modulation classical and quantum dispersions deviate slightly both in position and in 
momentum space. As we increase the 
modulation in the system, momentum and position dispersions show contrasting behavior in classical and in 
quantum mechanical domain. In classical dynamics dispersion both in position space and in momentum space 
increases while, quantum dispersion displays saturation. 
The quantum dispersion manifests periodic oscillations in momentum space and in position space. 
At the minima of these oscillations we observe strong localization at particular modulation values 
where the difference between classical dispersion and quantum mechanical dispersion is maximum.  
As the value of rescaled Planck's constant, $k^{\hspace{-2.1mm}-}$ reduces, the time evolution of the 
quantum dispersion both in position 
and momentum space approaches to classical dispersion and periodic oscillation in the quantum mechanical 
dispersion vanish as well, hence, the periodic behavior is strictly quantum phenomena which may display 
quantum revivals in dynamical systems. 

\section {Conclusion}\label{sec:Discussion}

In this contribution we study the dynamics of a condensed atoms in an opto-mechanical system where it is 
coupled to the moving end mirror
through the cavity field and displays dynamically localization both in position and momentum space. 
A contrast between classical and quantum dynamics is the evidence of the existence of dynamical localization 
and its presence is the signature of quantum chaos in a dynamical system. This behavior is different from 
the dynamical localization of ultra-cold atoms in modulated optical field where localization is observed only in 
momentum space \cite{Moore1994}. 
Spatio-temporal dynamics of condensate in position and momentum space confirm our finding as well. 
We have also calculated the dispersion in position and momentum space as a function of modulation 
for fixed time.
The dynamical localization phenomenon is realizable  
experimentally in position and in momentum space by using presently available experimental parameters.

\end{document}